\documentclass[%
reprint,
amsmath,amssymb,
aps,superscriptaddress
]{revtex4-2}

\usepackage{graphicx} 
\usepackage{dcolumn} 
\usepackage{bm} 
\usepackage{hyperref}
\usepackage{tabularx} 
\usepackage{siunitx} 

\hypersetup{
	colorlinks   = true, 
	urlcolor     = blue, 
	linkcolor    = blue, 
	citecolor    = blue 
}

\usepackage{soul}

\begin{document}
	
\preprint{APS/123-QED}

\title{Optical N-plasmon: Topological hydrodynamic excitations in Graphene from repulsive Hall viscosity}

\author{Wenbo Sun}
 \altaffiliation{equal contribution}
 \affiliation{Elmore Family School of Electrical and Computer Engineering, Birck Nanotechnology Center, Purdue University, West Lafayette, Indiana 47907, USA}
\author{Todd Van Mechelen}
 \altaffiliation{equal contribution}
 \affiliation{Elmore Family School of Electrical and Computer Engineering, Birck Nanotechnology Center, Purdue University, West Lafayette, Indiana 47907, USA}
 \author{Sathwik Bharadwaj}
 \affiliation{Elmore Family School of Electrical and Computer Engineering, Birck Nanotechnology Center, Purdue University, West Lafayette, Indiana 47907, USA}
 \author{Ashwin K. Boddeti}
  \affiliation{Elmore Family School of Electrical and Computer Engineering, Birck Nanotechnology Center, Purdue University, West Lafayette, Indiana 47907, USA}
 \author{Zubin Jacob}
 \email{zjacob@purdue.edu}
 \affiliation{Elmore Family School of Electrical and Computer Engineering, Birck Nanotechnology Center, Purdue University, West Lafayette, Indiana 47907, USA}

\begin{abstract}
Edge states occurring in Chern and quantum spin-Hall phases are signatures of the topological electronic band structure in two-dimensional (2D) materials. Recently, a new topological electromagnetic phase of graphene characterized by the optical N-invariant has been proposed. Optical N-invariant arises from repulsive Hall viscosity in hydrodynamic many-body electron systems, fundamentally different from the Chern and $Z_2$ invariants. In this paper, we introduce the topologically protected edge excitation -- optical N-plasmon of interacting many-body electron systems in the topological optical N-phase. These optical N-plasmons are signatures of the topological plasmonic band structure in 2D materials.  We demonstrate that optical N-plasmons exhibit fundamentally different dispersion relations, stability, and edge profiles from the topologically trivial edge magneto plasmons. Based on the optical N-plasmon, we design an ultra sub-wavelength broadband topological hydrodynamic circulator, which is a chiral quantum radio-frequency circuit component crucial for information routing and interfacing quantum-classical computing systems. Furthermore, we reveal that optical N-plasmons can be effectively tuned by the neighboring dielectric environment without breaking the topological properties. Our work provides a smoking gun signature of repulsive Hall viscosity and opens practical applications of topological electromagnetic phases of two-dimensional materials.
\end{abstract}

\maketitle

\section{Introduction}\label{section1}
Over the past few decades, the discoveries of topological phases and protected edge excitations of two-dimensional materials have gained a prominent role in condensed matter physics and photonics~\cite{bernevig2006quantum,rechtsman2013photonic,lu2014topological}. In graphene, the Chern invariant $(C \in Z)$ originating from complex electron next-nearest-neighbor (NNN) hopping was first proposed to achieve a topological electronic phase without external magnetic fields~\cite{haldane1988model}. The study of the corresponding chiral edge charge transport inspired discoveries beyond condensed matter physics ~\cite{hao2008topological}, in photonics~\cite{hafezi2013imaging,poo2011experimental}, cold atoms~\cite{jotzu2014experimental}, and acoustics~\cite{yang2015topological}. On the other hand, the $Z_2$ invariant $(\nu \in Z_2)$ emerges in graphene in the presence of spin-orbit coupling and characterizes the quantum spin Hall phase~\cite{kane2005quantum}. Insights into the associated chiral edge spin transport have driven potential applications in spintronics~\cite{konig2008quantum,brune2012spin,han2018quantum} and topological light sources~\cite{harari2018topological,xie2020higher}. 

Recently, a new topological electromagnetic phase of graphene characterized by the optical N-invariant $(N \in Z)$ was proposed~\cite{van2021optical,van2022optical}. This new topological phase arises only in the hydrodynamic regime of the interacting many-body electron system. The optical N-invariant characterizes the topology of bulk plasmonic  as opposed to the electronic band structure and arises from the Hall viscosity of electron fluids. It is fundamentally different from the Chern and $Z_2$ invariant characterizing the topology of bulk electronic bands in graphene~\cite{van2021optical}. Inspired by this development, in this article, we introduce the topologically protected edge state -- optical N-plasmon of this topological optical N-insulator and explore potential applications as well as control techniques.

Recent interest has focused on the hydrodynamic regime of graphene in the electronic context, such as the violation of the Wiedemann-Franz law~\cite{principi2015violation,ahn2022hydrodynamics,crossno2016observation} and negative local resistance~\cite{bandurin2016negative,levitov2016electron}. However, we note that the unique topological plasmonic behavior in the hydrodynamic regime is relatively unexplored. Our article here combines electrodynamics and hydrodynamics of graphene to uncover the topological properties. A related quantity, Hall viscosity in the static regime was measured for the first time recently~\cite{berdyugin2019measuring} even though the theoretical prediction was made two decades ago~\cite{avron1998odd}. It was shown that $\nu_H$ is connected to non-local Hall conductivity, non-local gyrotropy, and topological acoustic waves~\cite{hoyos2012hall,sherafati2016hall,van2018quantum,van2019photonic,van2019nonlocal,souslov2019topological,tauber2020anomalous,rapoport2023generalized}.

\begin{figure*}[t]
    \centering
    \includegraphics[width = 5.6in]{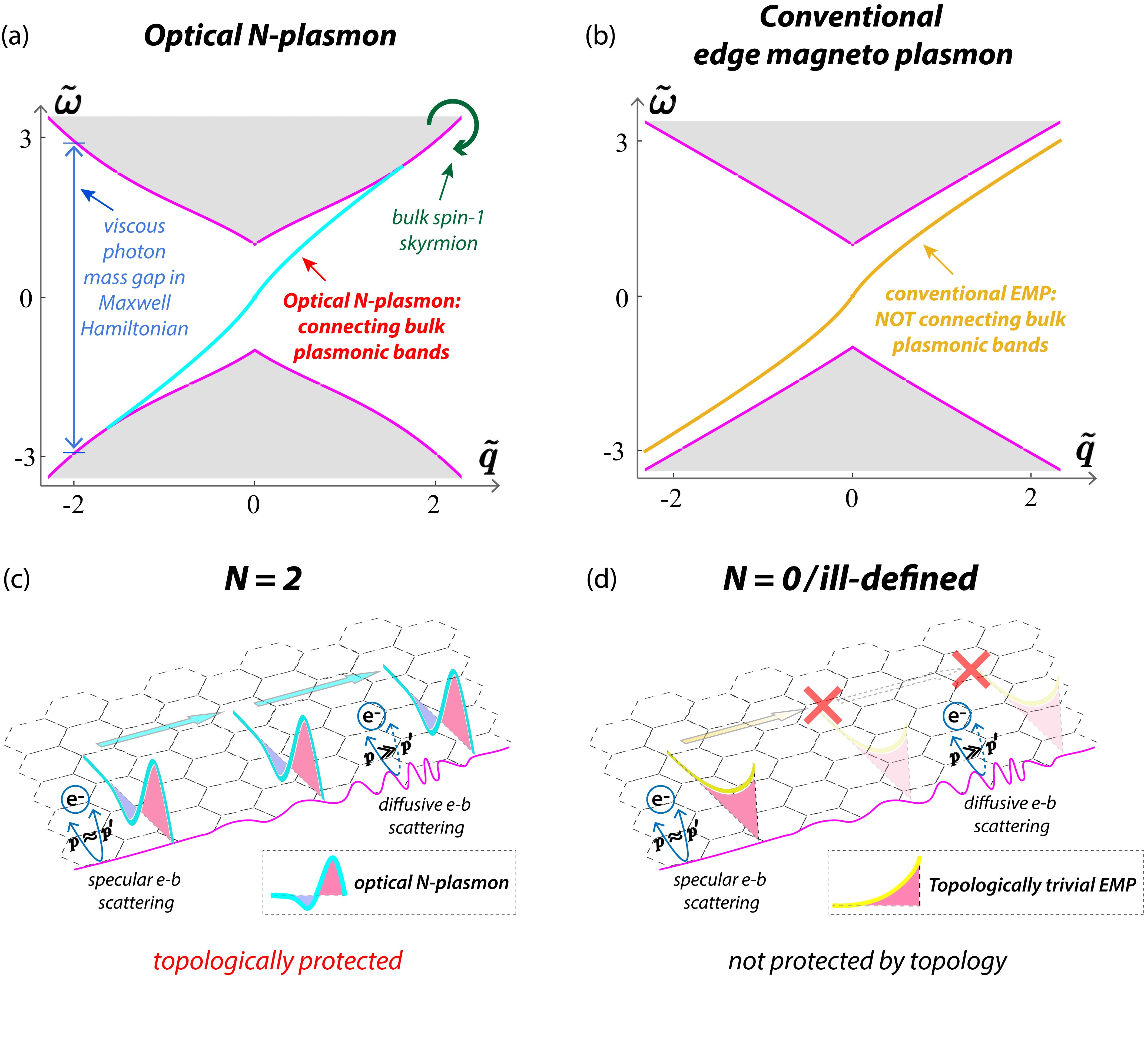}
    \caption{Topologically protected optical N-plasmons are fundamentally different from conventional edge magneto plasmons (EMPs). (a) The dispersion of optical N-plasmons (cyan) and bulk magnetoplasmons (magenta) in two-dimensional (2D) electron fluid in the topological electromagnetic (EM) phase. Optical N-plasmon is the edge signature of the topological EM phase. Bulk signatures of the topological EM phase include the viscous photon mass gap and bulk spin-1 skyrmion~\cite{van2020viscous}. The dispersion of the optical N-plasmon connects bulk plasmonic bands. (b) The dispersion of conventional EMPs~\cite{fetter1985edge} (yellow) and bulk magnetoplasmons (magenta) in 2D electron systems \emph{not} in the topological EM phase. The dispersion of conventional EMP fails to connect bulk plasmonic bands. (c) Optical N-plasmons are immune to back-scattering at boundary defects and not sensitive to varying degrees of edge disorders. Edge disorders can change electron-boundary scattering drastically from specular type, where only a small portion of electron momentum is transferred to the boundary (electron momentum $p$ before scattering $ \approx p'$ after scattering), to diffusive type, where nearly all electron momentum is lost during the scattering ($p \gg p'$). (d) EMPs not protected by topology can suffer from back-scattering and are not stable in the presence of varying degrees of edge disorders.}
    \label{fig:fig1}
\end{figure*}

In this paper, we introduce the optical N-plasmon, a unique topological edge excitation that only occcurs in the many-body interacting hydrodynamic regime of graphene. We demonstrate that optical N-plasmons are fundamentally different from topologically-trivial edge magneto plasmons (EMPs), including the conventional EMP~\cite{fetter1985edge} in the characteristic dispersion relations, stability with respect to edge disorders, and edge profiles. We show that the dispersion of optical N-plasmons exhibits nontrivial topological nature and closes the bulk plasmonic bandgap (Fig.~\ref{fig:fig1}(a)). In stark contrast, dispersions of topologically-trivial EMPs fail to do so in general (Fig.~\ref{fig:fig1}(b)). We further reveal that, since the optical N-plasmon is topologically protected, it is not sensitive to either sharp boundary defects or edge disorders that can change the nature of electron-boundary scattering properties from diffusive to specular (see Fig.~\ref{fig:fig1}(c)). In contrast, EMPs not protected by topology can suffer from back-scattering and are generally unstable when certain edge disorders are present (see Fig.~\ref{fig:fig1}(d)). Finally, we also discuss that optical N-plasmons provide the experimental smoking gun signatures of the optical N-invariant in the 2D electron fluid. 

Our study provides a rigorous comparison of different regimes for the emergence of optical N-plasmons and other plasmonic excitations in graphene. Graphene provides an important platform for studying plasmonic excitations in the 2D interacting many-body electron system in different regimes~\cite{koppens2011graphene,jablan2009plasmonics,garcia2014graphene,chen2012optical,grigorenko2012graphene,zhao2023observation}. In the non-interacting 2D electron gas (2DEG) regime, conventional gapless graphene plasmons and gapped graphene magneto plasmons were studied by identifying the zeros of dielectric functions~\cite{jablan2013plasmons,hwang2007dielectric,yan2012infrared,berman2008magnetoplasmons}. We notice that non-local effects on conventional graphene (magneto) plasmons are considered within the random phase approximation~\cite{hwang2007dielectric}. Conventional EMPs also emerge in the non/weakly interacting regime where the 2D electron system can be described by the Euler equation without any viscous term~\cite{fetter1985edge}. Meanwhile, optical N-plasmons proposed in this article emerge only in the strongly-interacting hydrodynamic flow regime with repulsive Hall viscosity. We obtain dispersion relations of optical N-plasmons by finding the propagating solutions of the underlying hydrodynamic equations. Therefore, non-local effects originating from the viscous hydrodynamic model are naturally included.  We develop an electromagnetic-hydrodynamic simulation based on the multiphysics model combining the linearized Navier-Stokes equations and electromagnetic equations. We employ experimentally-relevant parameters for simulating the optical N-plasmons in the 2D graphene interacting many-body electron system. 

Based on the optical N-plasmons, we propose the design of an ultra sub-wavelength broadband topological hydrodynamic circulator. Circulators are non-reciprocal circuit components important for microwave communications and quantum-classical information routing~\cite{stace2004mesoscopic,kerckhoff2015chip,chapman2017widely}. 
Many conventional ferrite or plasmonic circulator designs~\cite{pardavi2000microwave,dmitriev2019ultrawideband,viola2014hall,mahoney2017chip} are based on chiral EMPs not protected by topology~\cite{volkov1985theory}. The topological hydrodynamic circulator inherits robustness from optical N-plasmons, and the circulation behavior will not be perturbed by boundary defects or edge disorders. We simulate the performance of the proposed topological circulator with realistic graphene parameters. We show that the simulated frequency, momentum, and edge profile of the optical N-plasmon match well with the topological theory. 

We reveal that the optical N-plasmon can be effectively tuned by the neighboring dielectric environment without breaking its topological properties. 
Engineering plasmon properties is crucial for manipulating light in nano-devices~\cite{alonso2014controlling}. We study the properties of optical N-plasmons in both transparent and opaque neighboring dielectric environments. We show that without introducing electrical contacts or structure deformations~\cite{kamata2010voltage,jin2017infrared,jiang2018group}, group velocities of optical N-plasmons can be tuned in a contact-free manner by controlling the fringing fields in neighboring dielectric materials. The controllability and the aforementioned compact and topological nature indicate potential applications of the optical N-plasmons in graphene plasmonics~\cite{koppens2011graphene,jablan2009plasmonics,jablan2013plasmons,garcia2014graphene,chen2012optical,grigorenko2012graphene,zhao2023observation}.

The paper is organized as follows. In Sec.~\ref{Ninvariant}, we discuss the hydrodynamic electron flow model and optical N-invariant. In Sec.~\ref{section2}, we study the dispersions and profiles of optical N-plasmons and other bulk and edge excitations in hydrodynamic electron fluids. We demonstrate the fundamental differences between the optical N-plasmon and other topologically trivial EMPs. In Sec.~\ref{section3}, we present the circulation of optical N-plasmons in the hydrodynamic topological circulator based on graphene electron fluids. In Sec.~\ref{section4}, we study the properties of optical N-plasmons in different neighboring dielectric environments. Section~\ref{section5} summarizes the paper and indicates further applications of optical N-plasmons for future research.

\section{Optical N-invariant}\label{Ninvariant}

For completeness, we first summarize some key aspects of the topological optical N-invariant in graphene's viscous Hall fluid. Interacting many-body electron systems in various two-dimensional (2D) materials can be described by the hydrodynamic electron flow model when the momentum-conserving electron-electron scattering is dominant~\cite{jenkins2022imaging,sulpizio2019visualizing,berdyugin2019measuring,lucas2018hydrodynamics,bandurin2016negative,pusep2022diffusion,moll2016evidence,gooth2018thermal,pellegrino2017nonlocal}. The optical N-invariant classifies the electromagnetic topology in the presence of electron-electron interactions through the bulk atomistic susceptibility tensor. It was shown that the optical N-invariant is the winding number of the atomistic susceptibility tensor~\cite{van2022optical}.  This response function tensor is a many-body Green's function of the system, which has both spatial and temporal dispersion (i.e., momentum and frequency dependence). Here, due to the \emph{f}-sum rule~\cite{gusynin2007sum} and Hall viscosity $\nu_H$, the susceptibility tensor is properly regularized. As a result, the originally unbounded 2+1D momentum-frequency space of this continuum model can be compactified and is topologically equivalent to $S^2\times S^1$~\cite{van2021optical,souslov2019topological}. Through the Green's function formalism~\cite{volovik2003universe,gurarie2011single}, a quantized integer topological invariant -- optical N-invariant can be defined for this interacting many-body system~\cite{van2021optical}:
\begin{equation}
    N=\mathrm{sgn}(\omega_c)+\mathrm{sgn}(\nu_H),
\end{equation}
where $\omega_c$ is the cyclotron frequency and $\nu_H$ is the Hall viscosity. The topological phase is characterized by $N=\pm 2$ in the presence of a repulsive Hall viscosity $\omega_c \, \nu_H >0$, and the topologically trivial phase is characterized by $N=0$ with $\omega_c \, \nu_H < 0$. The optical N-invariant represents the topological property of the bulk plasmonic band structure and is fundamentally different from the Chern invariant and $Z_2$ invariant that are related to the bulk electronic band structure~\cite{haldane1988model,kane2005quantum}. 

We emphasize that the topological hydrodynamic excitations in this paper have important implications beyond the linearized continuum model and can be generalized to include the lattice symmetry and local-field effects~\cite{van2022optical}. The topological protection is robust beyond the linear regime. The proof is related to the recently developed viscous Maxwell-Chern-Simmons theory, which connects the optical N-invariant with spin-1 eigenvalues at high-symmetry points~\cite{van2020viscous} in momentum space. The U(1) gauge field of the 2D interacting fluid has a twist captured by the flip of spin-1 eigenvalues at high symmetry points. Thus any impurity or perturbation which does not cause spin-flipping for ultra-subwavelength (high momentum) plasmonic waves will not open the bandgap (between edge and bulk plasmonic states) in a topological optical-N insulator. 

The optical N-plasmon introduced in this paper occurs on the edge and is a smoking gun signature of repulsive Hall viscosity. For the bulk magneto-plasmons in hydrodynamic graphene, there exists a spin-1 skyrmionic behavior in momentum space~\cite{van2019photonic}. The experimental probe of this momentum space skyrmion was predicted to be evanescent magneto-optic Kerr effect (e-MOKE) spectroscopy~\cite{van2021optical}. The sign change of the e-MOKE angle can shed light on this unique optical N-invariant of matter. We note that the Chern invariant and Z2 invariant of graphene do not capture these effects arising only in the many-body interacting hydrodynamic regime.

Finally, we note that these unique edge states occur as super-symmetric partners between spin-1 excitations in Maxwell's equations and spin-1/2 fermions in the Dirac equation~\cite{van2018quantum}. Gyrotropy in Maxwell's equations is analogous to an effective photon mass when compared to mass in the 2D Dirac equation~\cite{van2020viscous,van2019unidirectional}. The topological edge excitations in Maxwell's equations only occur from dispersive photon mass (non-local gyrotropy) of a specific sign: i.e., repulsive Hall viscosity. On the other hand, attractive Hall viscosity leads to a topologically trivial phase. Thus the signature of an optical N-phase in an ideal model can be considered to be massive spin-1 excitations in the bulk and massless linearly dispersing spin-1 excitations on the edge.  However, no candidate material was known to exhibit these effects. One of our aims is to prove that graphene can exhibit these unique effects for experimental exploration.
\section{Optical N-plasmons}\label{section2}

\subsection{Hydrodynamic electron flow model}

In this part, we present the hydrodynamic electron flow model considered in this work.
In the hydrodynamic regime, electron transport is governed by the linearized Navier-Stokes equation with a viscous term~\cite{torre2015nonlocal,fetter1985edge,li2022hydrodynamic,levitov2016electron}. The anti-symmetric part of the viscous tensor, Hall viscosity $\nu_H$, can emerge in the 2D electron fluid when both time-reversal and parity symmetries are broken by an external magnetic field~\cite{avron1995viscosity,bradlyn2012kubo,avron1998odd,read2009non,gusev2018viscous,burmistrov2019dissipative,afanasiev2022hall,alekseev2016negative,sherafati2016hall,rao2023resolving,holder2019unified,narozhny2019magnetohydrodynamics,varnavides2020electron}. This non-dissipative Hall viscosity $\nu_H$ was first measured in ultra-clean graphene~\cite{berdyugin2019measuring}. For an interacting many-body electron system, when the momentum-conserving electron-electron scattering is dominant, the linearized Navier-Stokes equations describing the hydrodynamics of electrons in 2D is~\cite{van2021optical,fetter1985edge,torre2015nonlocal}:
\begin{subequations}\label{equation1}
\begin{multline}\label{equation1a}
   \frac{\partial \mathbf{J}}{\partial t}=-v_s^2\nabla \rho - (\gamma-\nu \mathbf{\nabla}^2) \mathbf{J}\\ 
   - (\omega_c+\nu_H \mathbf{\nabla}^2) \mathbf{J} \times \hat{z} +\frac{e^2n_0}{m}\mathbf{E},
\end{multline}
\begin{equation}\label{equation1b}
    \partial_t\rho+\pmb{\nabla}\cdot\mathbf{J}=0.
\end{equation}
\end{subequations}

Here, $\mathbf{J}=(J_x,J_y)$ is the 2D current density, $v_s^2=v_F^2/2$ represents the compressional wave velocity in the 2D electron fluid, $v_F$ is the Fermi velocity, $\rho$ is the charge density, $\gamma$ is the damping rate, $\nu$ is the ordinary shear viscosity, $\omega_c=eB/(mc)$ is the cyclotron frequency, $\nu_H$ is the Hall viscosity, $n_0$ is the electron density, and $m$ is the effective electron mass. Within the quasi-static approximation, electric field $\mathbf{E}=-\nabla \phi$. In the absence of external free charges out of the electron fluid plane, $\mathbf{E}$ arises from the fringing fields in the surrounding medium. The second term on the RHS of Eq.~(\ref{equation1a}) describes dissipation in the electron fluid. Meanwhile, the third term $(\omega_c+\nu_H \mathbf{\nabla}^2) \mathbf{J} \times \hat{z}$ is dissipation-less and will only emerge in the 2D electron fluid when both time reversal symmetry $\mathcal{T}$ and parity symmetry $\mathcal{P}$ are broken at the same time by an external magnetic field $B$. Continuity Eq. (\ref{equation1b}) describes the charge conservation law for electrons.

\subsection{Bulk magneto plasmons}

\begin{figure*}
    \centering
    \includegraphics[width = 6.8in]{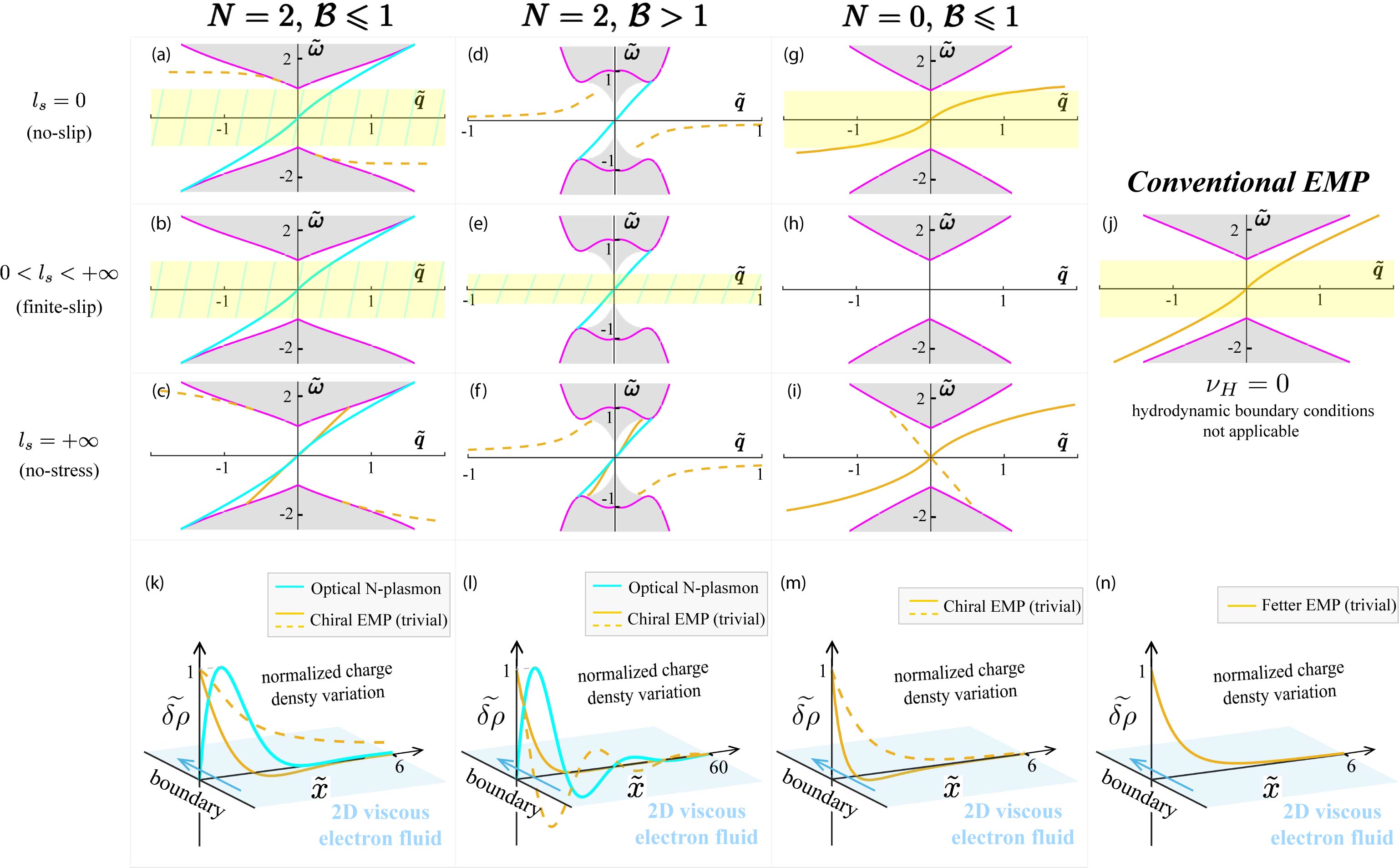}
    \caption{Classification of different bulk and edge excitations in hydrodynamic electron fluids. Dispersions of topologically protected optical N-plasmons, topologically trivial edge magneto plasmons (EMPs), and bulk plasmons are represented by cyan, yellow, and magenta curves, respectively. Gray regions represent scattering bulk plasmonic modes. (a-f) Dispersions of optical N-plasmons and other EMPs when the bulk is in the $N=2$ phase with $B \leqslant 1$ (a-c) or $B > 1$ (d-f) under different electron fluid boundary conditions characterized by slip length $l_s=0$ (a,d), $0<l_s<+\infty$ (b,e), and $l_s=+\infty$ (c,f). Topologically protected optical N-plasmons are not sensitive to $l_s$ and connect bulk bands in all cases. Other EMPs can only exist under certain $l_s$ and do not have topological origins. (g-i) Dispersions of EMPs when the bulk is in the $N=0$ phase. Optical N-plasmons do not exist because the bulk is topologically trivial. (j) Dispersions of conventional Fetter EMPs when Hall viscosity $\nu_H$ is absent in the electron fluid. In this case, optical N-invariant is ill-defined, and there is no topological interpretation for bulk plasmonic modes. As a result, the conventional Fetter EMPs are not protected by topology and can not connect bulk bands. (k-n) Edge profiles $\widetilde{\delta \rho}(\tilde{x})$ of optical N-plasmons (cyan) and chiral/Fetter EMPs (yellow). $\widetilde{\delta \rho}(\tilde{x})$ represents the normalized charge density variations at distance $\tilde{x}$ away from material boundaries. Optical N-plasmons have distinctively different edge profiles from chiral/Fetter EMPs. The plots indicate the frequency windows for unidirectional edge excitations (yellow) and topologically protected edge excitations -- optical N-plasmons (yellow-cyan).}
    \label{fig:fig2}
\end{figure*}

We first discuss the bulk magneto plasmons in the 2D electron fluid with Hall viscosity. We consider the dielectric material surrounding the 2D electron fluid is isotropic with an effective permittivity tensor $\overleftrightarrow{\varepsilon}=\varepsilon \overleftrightarrow{I}$. From Eq.~(\ref{equation1a},\ref{equation1b}), in the low-loss limit ($\gamma,\nu \rightarrow 0$), we can solve the dispersion of bulk magneto plasmons by considering propagating bulk modes of the form $e^{i(\mathbf{q} \cdot \mathbf{r} - \omega t)}$:
\begin{equation}\label{bulk}
    \omega^2=\frac{2\pi e^2 n_0 |q|}{m \varepsilon}+v_s^2 q^2 +(\omega_c - \nu_H q^2)^2.
\end{equation}

Here, the last term in Eq.~(\ref{bulk}) originates from the external magnetic field and opens the bandgap between bulk bands at $q=0$. As $q\rightarrow + \infty$, the bulk magneto plasmon dispersion $\omega(q)$ is dominated by the Hall viscosity term and shows the asymptotic behavior $\omega=\mathcal{O}(q^2)$. In contrast, in the absence of $\nu_H$, the conventional bulk magneto plasmon dispersion is dominated by the $v_s^2 q^2$ term with $\omega=\mathcal{O}(q)$ when $q\rightarrow + \infty$. We focus on the transparent surrounding material with $\varepsilon>0$. Hence, the bandgap between bulk bands will be opened for all momentum $q$. The shape of the bulk magneto plasmon dispersion largely depends on the Hall viscosity $\nu_H$ and dielectric permittivity of the surrounding medium $\varepsilon$. We can define a unitless value $\mathcal{B}$ to classify two different shapes of bulk bands:
\begin{equation}
    \mathcal{B}=\frac{(2 \nu_H \omega_c - v_s^2)^3 m^2 \varepsilon^2 }{27 \pi^2 e^4 \nu_H^2 n_0^2}.
\end{equation}

For $\mathcal{B} \leqslant 1$, bulk bands will monotonically increase with momentum $|q|$. For $\mathcal{B}>1$, bulk bands will have a Mexican hat shape. In Fig.~\ref{fig:fig2}, we show these two classes of bulk bands by magenta curves (Fig.~\ref{fig:fig2}(a-c,g-i) for $\mathcal{B} \leqslant 1$, Fig.~\ref{fig:fig2}(d-f) for $\mathcal{B} > 1$). In the figures, $\tilde{q}=q \, v_s/\omega_c$ and $\tilde{\omega}=\omega/\omega_c$ are the unitless momentum and frequency normalized by characteristic parameters of the system. It is worth noting that for $\mathcal{B} \leqslant 1$, the bandgap of bulk bands $\Delta=2\omega_c$ is only determined by the external magnetic field $B$. For $\mathcal{B}>1$, the bandgap of bulk bands $\Delta<2\omega_c$ is also controlled by material properties and permittivity of the surrounding dielectric material.

\subsection{Optical N-plasmons}

In this part, we demonstrate the nontrivial topological properties of optical N-plasmons. We compare the dispersions of optical N-plasmons and other topologically trivial edge states in 2D electron fluid. Conventional edge states are usually believed to depend on boundary conditions sensitively~\cite{kiselev2019boundary}. In contrast, we show that the topologically protected optical N-plasmons are not sensitive to boundary conditions at the 2D electron fluid boundaries.

For the 2D electron fluid, fringing fields in the surrounding media and electron fluid boundary conditions complicate the edge problems significantly~\cite{fetter1985edge}. The fringing fields mediate the interactions between quasi-static charges in the 2D plane and contribute to an effectively non-local potential. We solve the edge problem with the non-local potential fully by numerical simulations in section~\ref{section3}. In this section, we adopt the Fetter approximation, which can provide accurate dispersions of edge states except in the long-wavelength limit ($q \rightarrow 0$)~\cite{fetter1985edge,cohen2018hall}. Boundary conditions of the 2D electron fluid are microscopically determined by degrees of edge disorders and the mechanism of electron-boundary scattering, and can be characterized by the slip length $l_s$~\cite{kiselev2019boundary,raichev2022linking}. Many different factors, including charge density $n_0$, temperature, and smoothness of material boundaries, can influence $l_s$. In the low-loss limit, electron fluid boundary conditions can be written in terms of $l_s$~\cite{kiselev2019boundary,pellegrino2017nonlocal}:

\begin{subequations}
\begin{equation}\label{bc}
\left[\hat{t}\cdot\varsigma\cdot\hat{n}+\hat{t}\cdot\mathbf{J}\right/l_s]=0,
\end{equation}

\begin{equation}
    \varsigma=\begin{bmatrix}
        -\partial_x J_y - \partial_y J_x & \partial_x J_x -\partial_y J_y \\
        \partial_x J_x -\partial_y J_y & \partial_x J_y + \partial_y J_x 
    \end{bmatrix},
\end{equation}
\end{subequations}
where $\hat{t}, \hat{n}$ are the unit vectors in the tangential and normal directions. The two extreme cases of electron fluid boundary conditions, no-slip and no-stress, correspond to slip length $l_s = 0$ and $l_s = \infty$, respectively. These two regimes could happen for the viscous electron fluid when the electron-boundary scattering is diffusive (no-slip) or specular (no-stress). In the intermediate regime where $0<l_s<\infty$, the finite-slip boundary condition is appropriate, where part of electron momentum is lost in the electron-boundary scattering process. 

In Fig.~\ref{fig:fig2}, we show the dispersions of edge excitations when the viscous Hall electron fluid is in different topological phases and under various boundary conditions. The derivations of the edge state dispersion and the material parameters are given in Appendix~\ref{appendixA}. The bulk-boundary correspondence guarantees that for the topological phase characterized by $N=2$, optical N-plasmons always exist in the bandgap of bulk bands. Dispersion of optical N-plasmons is marked by cyan curves in Fig.~\ref{fig:fig2}(a\,-\,c) for bulk bands with $\mathcal{B} \leqslant 1$ and in Fig.~\ref{fig:fig2}(d\,-\,f) for bulk bands with $\mathcal{B} > 1$. Optical N-plasmons can connect the bulk bands in both cases. Conventional edge states are usually considered to be sensitive to boundary conditions~\cite{kiselev2019boundary}. In contrast, as is shown in Fig.~\ref{fig:fig2}(a\,-\,f), the dispersion of the optical N-plasmon is independent of the fluid boundary conditions since it is protected by topology. This reveals the advantages of optical N-plasmons for practical applications in information technology, where stability is highly required. As a result, we simulate the performance of a circulator in section \ref{section3} based on the optical N-plasmons. Apart from the optical N-plasmon, some other types of edge magneto plasmons (EMPs) can also exist in the bandgap under extreme boundary conditions ($l_s=0 \ \mathrm{or} \ \infty$). These EMPs may also be chiral (CEMPs) and are marked by yellow solid and dashed curves in Fig.~\ref{fig:fig2}. It is worth noting that although optical N-plasmons are always chiral, CEMPs are not necessarily protected by topology. CEMPs exist in the bandgap due to the anomalous bulk-boundary correspondence under specific boundary conditions. It is related to the scattering of bulk modes at the boundary and ``ghost edge modes'' at infinite frequency~\cite{tauber2020anomalous}. Dispersions of CEMPs are very sensitive to boundary conditions.  As is shown in Fig.~\ref{fig:fig2}(a,\,c,\,d,\,f), by continuously deforming the shape of bulk bands without closing the bandgap at any $\tilde{q}$ point, the group velocity of CEMP can be reversed, which is in contrast to the optical N-plasmon. The frequency windows where only optical N-plasmons can be excited are marked by the yellow-cyan regions.

In Fig.~\ref{fig:fig2}(g\,-\,i), we present the dispersions of edge states for the topologically trivial $N=0$ phase. Here, optical N-plasmons do not exist. Despite some unidirectional frequency windows existing under extreme cases of boundary conditions (marked by yellow region), the dispersions of these CEMPs are not stable under varying boundary conditions. Furthermore, the bandgap of bulk bands can not be connected under all boundary conditions. This is because the bulk material is in a topologically trivial phase, and no edge state is protected by topology. Figure.~\ref{fig:fig2}(j) shows the dispersion of the conventional Fetter edge magneto plasmons~\cite{fetter1985edge} (FEMP) for $\nu_H=0$. In this case, since the unbounded momentum space can not be compactified due to the absence of Hall viscosity $\nu_H$, no topological interpretation exists for the bulk. As a result, FEMP is unidirectional but not topological and can not connect bulk bands. Hence, FEMP is not guaranteed to be immune to back-scattering at the boundary defects.

\begin{figure*}
    \centering
    \includegraphics[width = 5.2 in]{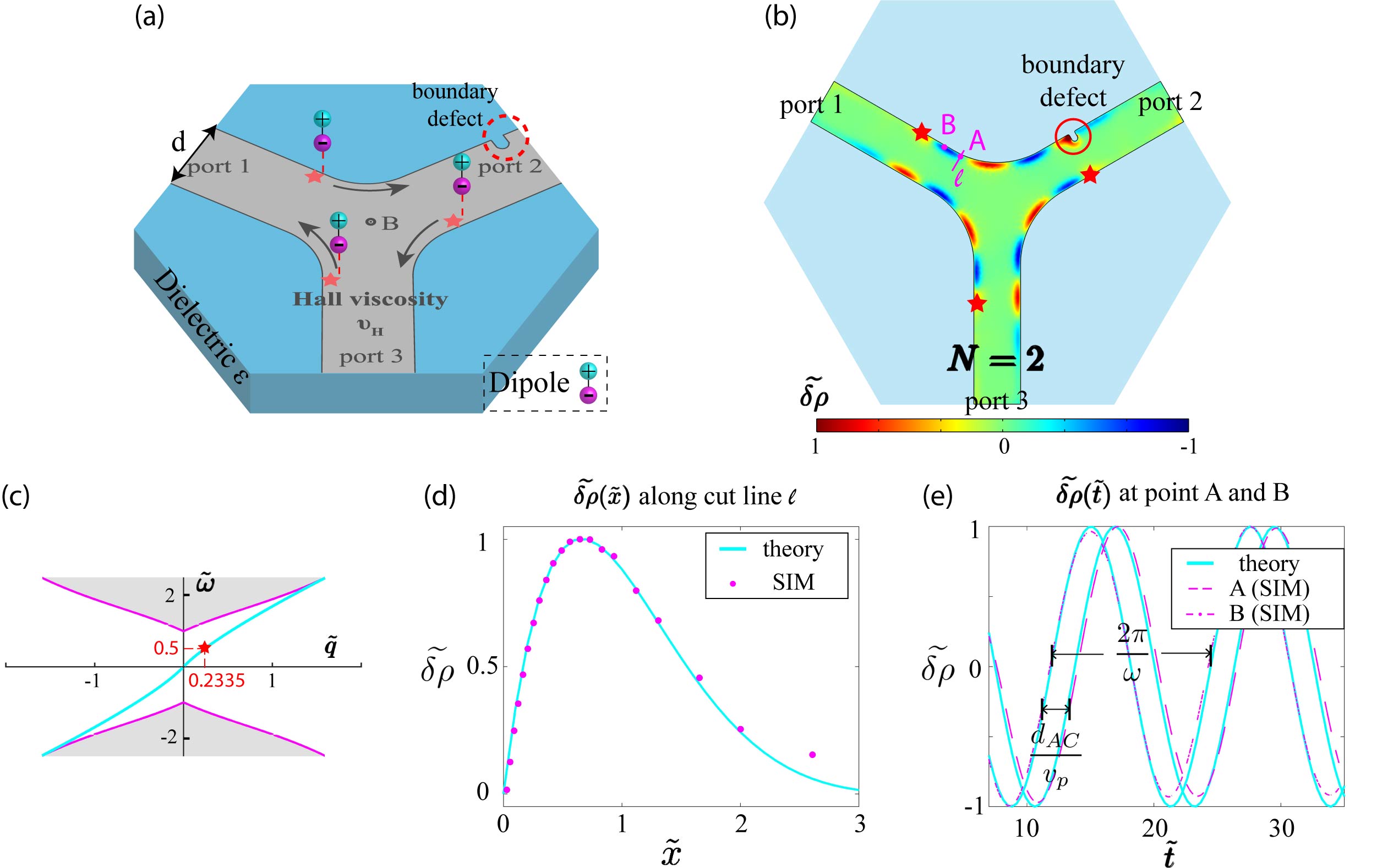}
    \caption{Ultra sub-wavelength broadband topological hydrodynamic circulator. (a) Schematic of the 3-port topological hydrodynamic circulator. The circulator is based on the electron fluid with repulsive Hall viscosity in ultra-clean graphene (gray) on top of the dielectric medium (blue) with an external magnetic field. Three oscillating dipoles excite the optical N-plasmons in three ports. The boundary defect in port 2 is marked by the red circle.  (b) Time-domain simulations of the topological hydrodynamic circulator. $\widetilde{\delta \rho}$ is the normalized charge density variation. Circulation of optical N-plasmons is immune to back-scattering at the boundary defect. (c) Dispersion of optical N-plasmons (cyan curve) in graphene's viscous hydrodynamic electron fluid. The red star marks the optical N-plasmon excited by the oscillating dipoles in the frequency and momentum space. (d) Simulated (magenta dots) and theoretical (cyan curve) edge profiles of optical N-plasmons in the circulator. (e) Charge density variation at points A and B at different time $\tilde{t}$. The simulated frequency and propagation velocity of optical N-plasmons match their theoretical counterparts.}
    \label{fig:fig3}
\end{figure*}

As is shown in Fig.~\ref{fig:fig2}(k\,-\,n), optical N-plasmons (solid cyan curves) have distinct normal profiles compared with CEMP and FEMP (solid and dashed yellow curves). Here, $\widetilde{\delta \rho}$ represents the charge density variation of different types of normalized edge states $\tilde{\psi}$. $\tilde{x}=x\, \omega_c/v_s$ is the normalized unitless distance from the fluid boundary. For optical N-plasmons, despite the confinement being related to the Hall diffusion length $D_H=\sqrt{\nu_H/\omega_c}$ and may vary with Hall viscosity, $\widetilde{\delta \rho}(\tilde{x}=0)=0$ is always valid. For CEMP and FEMP, $\widetilde{\delta \rho}(\tilde{x}=0)\neq 0$. This difference is because the dispersion of optical N-plasmons is independent of while dispersions of CEMP and FEMP are sensitive to boundary conditions. With some algebra, we can prove that in the low-loss limit, $\widetilde{\delta \rho}(\tilde{x}=0)=0$ is a necessary and sufficient condition for $\tilde{\psi}$ to be independent of varying boundary conditions (see Appendix A). 

In the next section, we employ optical N-plasmons to design an ultra sub-wavelength broadband topological hydrodynamic circulator.

\section{Topological hydrodynamic circulator}\label{section3}

\subsection{Fringing fields and non-local in-plane potential}\label{subsection3_1}

For the 2D electron fluid confined in the $z=0$ plane, the fringing fields out of the plane introduce a non-local effect in the coupling between the charge density $\rho$ and in-plane potential $\phi$. The non-local coupling between $\phi$ and $\rho$ confined in the 2D domain $\Omega$ and free charges $\rho_f$ out of the plane is:
\begin{equation}\label{potential}
\phi(t,\mathbf{r})=\frac{4\pi}{\varepsilon} \int_{\Omega} d\mathbf{r}'G(\mathbf{r},\mathbf{r}')\rho(t,\mathbf{r}')+\phi_f(t,\mathbf{r}),
\end{equation}
where $\phi_f(t,\mathbf{r})=4\pi\int d\mathbf{R_0} G(\mathbf{r},\mathbf{R_0})\rho_f(t,\mathbf{R_0})/\varepsilon$ is the electric potential generated from the free free charges $\rho_f$, $\varepsilon$ is the effective permittivity of the surrounding medium, $\mathbf{r}, \mathbf{r'}$ denote the 2D coordinates, $\mathbf{R_0}$ denotes the 3D coordinates, $G$ is the scalar Green's function,
\begin{equation}\label{greenfunction}
G(\mathbf{r},\mathbf{r}')=\frac{1}{4\pi |\mathbf{r}-\mathbf{r}'|}.
\end{equation}

Here, in contrast to the 3D case, for the 2D electron fluid, no simple differential operator with respect to the 2D coordinates $\mathbf{r}$ can relate $\phi$ and $\rho$ locally~\cite{fetter1985edge}. In this section, we develop an electromagnetic-hydrodynamic simulation to solve the coupled Eqs.~(\ref{equation1}) and (\ref{potential}).

\subsection{Graphene-based topological hydrodynamic circulator}\label{subsection3_2}

Optical N-plasmons at the edge of the viscous Hall electron fluid in the $N=2$ phase are fundamentally protected by the topology and are robust against fluctuations. As a result, it is well suited for applications in information processing. In this section, we propose the design of an ultra sub-wavelength broadband topological hydrodynamic circulator based on the optical N-plasmons in graphene.

The schematic of the 3-port circulator design is demonstrated in Fig.~\ref{fig:fig3}(a). Graphene with the Y-shape circulator geometry (gray region) is on top of the isotropic dielectric material with permittivity $\varepsilon_b$ (blue bulk). In this case, effective permittivity $\varepsilon=\varepsilon_b/2$. Graphene is required to be ultra-clean so that the interacting electrons can be described by the hydrodynamic flow model. A static external magnetic field is applied in the graphene region, and Hall viscosity can emerge in the system since the time-reversal symmetry and parity symmetry are broken. For repulsive Hall viscosity ($\nu_H \omega_c >0$), viscous Hall electron fluid in graphene will be in the topological $N=2$ phase. Three oscillating electric dipoles with oscillation frequency $\omega_s$ are placed on top of each port. These dipoles are used to excite optical N-plasmons in the circulator. Hence, $\omega_s$ is considered to be in the bandgap of bulk bands. The possible boundary defects of the circulator are captured by a sharp corner in port 2. Since optical N-plasmons are unidirectional and immune to back-scattering, the topological circulation behavior from $\mathrm{port}\, 1 \rightarrow \, 2 \rightarrow \, 3$ will not be interfered with by the boundary defect. Reversing the direction of the magnetic field realizes the topological phase transition into $N=-2$, and the circulator will have an opposite circulation direction $\mathrm{port}\, 3 \rightarrow \, 2 \rightarrow \, 1$ accordingly.

We employ the finite element method to simulate the topological hydrodynamic circulator in the time domain and demonstrate the topological circulation behavior of optical N-plasmons in Fig.~\ref{fig:fig3}(b). We also provide a supplementary video generated from the electromagnetic-hydrodynamic simulations. The graphene region is described by Eqs.~(\ref{equation1},\ref{potential}) and a finite slip boundary condition $0<l_s<\infty$ is applied at the boundary of graphene. In the simulations, we employ experimental graphene parameters (Appendix.~\ref{simulaton_details}) in the low-loss limit with $\mathcal{B}<1$ and consider a high index substrate with $\varepsilon=50$ under graphene. An external magnetic field $B=2\, \mathrm{T}$ is applied in the graphene region with a port width of $329\, \mathrm{nm}$. The three dipoles on top of each port with oscillation frequencies $\omega_s=\omega_c/2$ contribute to $\phi_f$ in Eq.~(\ref{potential}). Their projections in the graphene plane are marked by red stars. Inside the graphene region, normalized charge density variations $\widetilde{\delta \rho}$ are represented by the colorbar. From the simulations, it is clear that the excited optical N-plasmons at red stars will flow unidirectionally from $\mathrm{port}\, 1 \rightarrow \, 2 \rightarrow \, 3$. Optical N-plasmons cross the sharp corner in port 2 smoothly without back-scattering.

In Fig.~\ref{fig:fig3}(c), we show the dispersion relation (cyan curve) of optical N-plasmons with the graphene parameters considered in our simulations. We mark the optical N-plasmons excited by the oscillating dipoles in our simulations with the red star in the frequency and momentum space. In Fig.~\ref{fig:fig3}(d), we show the normal profile of the normalized edge excitation at $\tilde{t}\approx23$ and compare the charge density variations along cut line segment $\ell$ in port 1 ($\ell$ is marked by the magenta line in Fig.~\ref{fig:fig3}(b)). $\tilde{x}$ represents the distance from the fluid boundary. $\tilde{t}=t \, \omega_c$ is the unitless time normalized by the characteristic timescale of the system. The simulation results in each mesh along $\ell$ match the theory predictions (cyan curve) well. The small deviations at larger $\tilde{x}$ are related to coarser meshes in that region. $\widetilde{\delta \rho}(\tilde{x}=0)=0$ shows that the excited edge states in Fig.~\ref{fig:fig3}(b) are optical N-plasmons protected by topology instead of other types of chiral edge states. In Fig.~\ref{fig:fig3}(e), we study $\widetilde{\delta \rho}(t)$ at points A and B and show the corresponding charge density variations (points A and B are marked by the magenta dots in Fig.~\ref{fig:fig3}(b)). The theory curves are plotted by fitting the simulation results at point A using a $sine$ curve with period $2\pi/\omega$ and translationally shifting it by $d_{AC}/v_p$ to get theory results at point B. We can see that the frequency $\omega$ and propagation velocity $v_p$ of the simulated optical N-plasmons match with their theoretical counterparts.

The performance of a circulator can be evaluated based on many aspects, including the form factor, isolation, and bandwidth. Form factor $\mathcal{F}$ indicates the relative size of a circulator with respect to its working frequency range. In our design, $\mathcal{F}$ is determined by the confinement of optical N-plasmons $D_H=\sqrt{\nu_H/\omega_c}$ and can be defined as the ratio between port width $d$ and vacuum wavelength corresponding to the frequency $\omega_s$. For the simulated circulator performance in Fig.~\ref{fig:fig3}(b), $\mathcal{F} \approx 2.5\times 10^{-3}$, revealing that this topological hydrodynamic circulator design is ultra-compact. Since no back-scattering is allowed by the topology and unidirectional optical N-plasmons are the only allowed state in the bandgap, this topological circulator should possess much larger isolation compared with other designs based on topologically-trivial edge states. The bandwidth of this topological circulator is determined by the bandgap of bulk bands. With an external magnetic field $B=2\, \mathrm{T}$, bandwidth $BW=\Delta=2\, \omega_c \approx 4.5 \,  \mathrm{THz}$ is ultrawide. It is worth noting that the performance of this design, including form factor, bandwidth, and response speed, can be effectively tuned and optimized by changing the external magnetic field or surrounding media (see section~\ref{section4}). All the discussions above show that the proposed design is ultra sub-wavelength, broadband, tunable, and can operate in the THz range. This reveals that the topological hydrodynamic circulator can play an important role in next-generation information routing and interfacing quantum-classical computing systems.

\section{Contact-free optical N-plasmon control with neighboring dielectric environment}\label{section4}

Although fringing fields in the surrounding medium can cause intrinsic non-locality in Eq.~(\ref{potential}) and complicate the problem greatly, they can offer new flexibility to tune optical N-plasmons. The existence of optical N-plasmons is guaranteed by topology regardless of the neighboring dielectric materials, but it is possible to exploit the surrounding medium to tune and optimize the optical N-plasmon properties in nano-devices without introducing electrical contacts in the viscous electron fluid. In this section, we study the influence of the surrounding dielectric environment on optical N-plasmons and the topological hydrodynamic circulator. 

\begin{figure}[t]
    \centering
    \includegraphics[width = 3.5in]{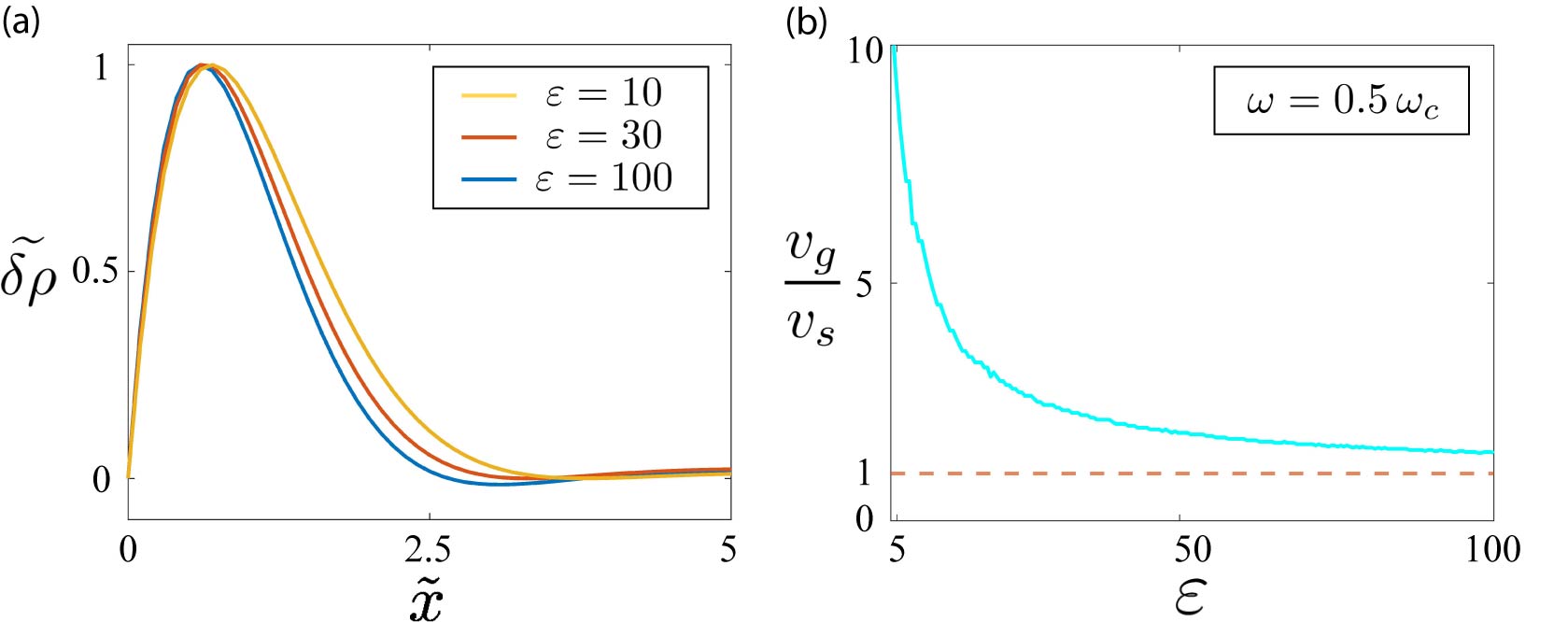}
    \caption{Effective control of optical N-plasmons by the dielectric medium with permittivity $\varepsilon$. (a) Confinement of optical N-plasmons at $\varepsilon=10,30,100$. Confinement is determined by Hall diffusion length and is not sensitive to $\varepsilon$. (b) The group velocity of optical N-plasmons at $5 \leqslant \varepsilon \leqslant 100$. Group velocity can be effectively tuned in a contact-free manner by surrounding medium permittivity $\varepsilon$.}
    \label{fig:fig4}
\end{figure}

We consider that the 2D graphene viscous electron fluid in the $N=2$ phase is on top of the isotropic transparent material with positive dielectric constant $\varepsilon$. The bandgap of bulk bands is always connected by optical N-plasmons at the edge. In Fig.~\ref{fig:fig4}(a), we show that the confinement of optical N-plasmons is not sensitive to $\varepsilon$. We demonstrate that the charge density variations corresponding to the normal profiles of the normalized optical N-plasmon states $\widetilde{\delta \rho}(\tilde{x})$ at $\varepsilon=10,\,30,\,100$ are similar. This is because the confinement of $\widetilde{\delta \rho}(\tilde{x})$ is determined by the Hall diffusion length $D_H=\sqrt{\nu_H/\omega_c}$ independent of $\varepsilon$. In Fig.~\ref{fig:fig4}(b), we present that the group velocity $v_g=v_s\,d\,\tilde{\omega}/d\,\tilde{q}$ of optical N-plasmons can be effectively tuned by $\varepsilon$. By changing $\varepsilon$ from $5$ to $100$, group velocity $v_g$ of the optical N-plasmon with $\tilde{\omega}=0.5$ can be modulated by a factor of 10. It is worth noting that $v_g > v_s=v_F/\sqrt{2}$ and can approach $v_s$ asymptotically in the large $\varepsilon$ limit. For the topological circulator design, the stable confinement of optical N-plasmons reveals that the circulator can always be ultra-compact, and the controllable $v_g$ indicates a tunable response speed of the topological circulator.

\section{Conclusion}\label{section5}
To summarize, we introduce the optical N-plasmon, which is the topologically protected edge excitation of the two-dimensional hydrodynamic electron flow with repulsive Hall viscosity. Optical N-plasmons are fundamentally different from conventional chiral/Fetter EMPs in three aspects: dispersion relations, stability with respect to edge disorders, and edge profiles. We propose an ultra sub-wavelength broadband topological hydrodynamic circulator based on optical N-plasmons, which is a chiral quantum radio-frequency circuit component crucial for information routing and interfacing quantum-classical computing systems. The topological circulator has a robust performance when boundary defects and edge disorders are present. The simulated optical N-plasmons circulating in the circulator ports show a good match with the theory. We demonstrate that group velocities of optical N-plasmons can be tuned in a contact-free manner by controlling the fringing fields in neighboring dielectric materials. Our work provides an experimental signature of repulsive Hall viscosity and opens practical applications of the new topological electromagnetic phase of two-dimensional materials. Moreover, the compact, tunable, topologically protected optical N-plasmons can have further applications in various fields, including graphene plasmonics~\cite{grigorenko2012graphene,koppens2011graphene,chen2012optical,jablan2013plasmons,jablan2009plasmonics,garcia2014graphene}, plasmonic metamaterials~\cite{monticone2017metamaterial}, nonreciprocal quantum devices~\cite{viola2014hall,bosco2017self}.

\section{Acknowledgements}
	This work was supported by the Defense Advanced Research Projects Agency (DARPA) under Nascent Light-Matter Interactions (NLM) program and U.S. Department of Energy (DOE), Office of Basic Sciences under DE-SC0017717.
 
\appendix

\section{Edge excitations of the hydrodynamic electron fluid}\label{appendixA}
In this appendix, we provide solutions to the edge excitations of the hydrodynamic electron flow model based on the Fetter approximation~\cite{fetter1985edge}.

Assuming the hydrodynamic electron fluid has the half plane geometry in the $x>0$ region. The propagating edge modes have the form $f(x,y,t)=f(x)e^{i(qy-\omega t)}$, where $f$ can be the charge density variation $\delta \rho$ or the 2D current density $\mathbf{J}$. As we have shown in Eq.~(\ref{potential}), the fringing fields out of the electron fluid plane introduce intrinsic non-locality in the electromagnetic potential $\phi$. Instead of solving this complex non-local problem for edge state dispersion, we consider an approximate integral kernel that makes the Poission's equation effectively local~\cite{fetter1985edge}. The non-local integral form of Poission's equation can thus be replaced by the differential equation:
\begin{equation}\label{fetter}
    \frac{\partial^2 \phi(x)}{\partial x^2} - 2q^2 \phi(x) = \frac{4\pi |q| \rho(x)}{\varepsilon}.
\end{equation}

This Fetter approximation provides accurate dispersion relations for our study except in the long-wavelength limit ($q \rightarrow 0$), where the asymptotic behavior of the exact solution can not be recovered~\cite{cohen2018hall}. Combining Eq.~(\ref{equation1}) and Eq.~(\ref{fetter}), the dispersions $\omega=\omega(q)$ of edge modes of the form $e^{i(qy-\omega t)}$ are given by the following coupled equations~\cite{van2021optical}:
\begin{subequations}
    \begin{multline}\label{eqa2a}
    -\alpha^4 \tilde{q'}^6+(2\alpha^2-1-\alpha^4 \tilde{q}^2)\tilde{q'}^4 + \\(\tilde{\omega}^2-\tilde{\omega_b}^2-\tilde{\Omega_p}^2+\alpha^4 \tilde{q}^4)\tilde{q'}^2
    + \tilde{q}^2 (\tilde{\omega}^2-1) = 0,
    \end{multline}
    \begin{equation}
        \det [F_{ij}]_{3 \times 3} = 0,
    \end{equation}
\end{subequations}
where $[F_{ij}]$ is a $3\times3$ matrix corresponding to boundary conditions:
\begin{subequations}
    \begin{equation}
        F_{1j}=\sqrt{2}|\tilde{q}| + \eta_j,
    \end{equation}
    \begin{equation}
        F_{2j}=[\tilde{\omega} \eta_j - (1-\alpha^2\tilde{q'_j}^2)\tilde{q}] \, (\tilde{q'_j}^2+\tilde{q}^2) \, / \, \tilde{q'_j}^2,
    \end{equation}
    \begin{equation}
        F_{3j}^0=[\tilde{\omega}\tilde{q}  - (1-\alpha^2\tilde{q'}_j^2)\eta_j] \,(\tilde{q'_j}^2+\tilde{q}^2) \, / \, \tilde{q'_j}^2,
    \end{equation}
    \begin{equation}
        F_{3j}^{+\infty}=\left( 2\tilde{q}[\tilde{\omega}\tilde{q}  - (1-\alpha^2\tilde{q'}_j^2)\eta_j] - \tilde{\omega}\tilde{q'}_j^2 \right) \, (\tilde{q'_j}^2+\tilde{q}^2) \, / \, \tilde{q'_j}^2,
    \end{equation}
\end{subequations}
where $\alpha=\sqrt{\omega_c\nu_H}/v_s$ is a unitless constant determined by the electron fluid. $\tilde{\omega}=\omega/\omega_c$ and $\tilde{q}=v_s \, q/\omega_c$ are the normalized frequency and momentum. Equation~(\ref{eqa2a}) is a cubic equation with respect to $\tilde{q'}^2$, and $\tilde{q'_i}^2$ is the $i\mathrm{th}$ root of Eq.~(\ref{eqa2a}). $\eta_i=\sqrt{\tilde{q}^2-\tilde{q'_i}^2}$ with $\mathrm{Re} \, \eta_i \geqslant 0$. Here, plasma frequency $\tilde{\Omega_p}=\beta \sqrt{|\tilde{q}|}$, bulk plasmonic band dispersion $\tilde{\omega_b}^2=(1-\alpha^2 \tilde{q}^2)^2+\tilde{q}^2+\beta^2|\tilde{q}|$ and $\beta=\sqrt{2\pi e^2 n_0/(m \varepsilon \omega_c v_s)}$. When the electron fluid boundary condition is no-slip with $l_s=0$ or no-stress with $l_s=+\infty$, $F_{3j}=F_{3j}^0$ or $F_{3j}=F_{3j}^{+\infty}$ respectively. For finite slip boundary condition, $F_{3j}=F_{3j}^{+\infty} + \kappa F_{3j}^{0}$, where $\kappa$ is a constant determined by $l_s$. 

\begin{figure}[t]
    \centering
    \includegraphics[width = 2.5 in]{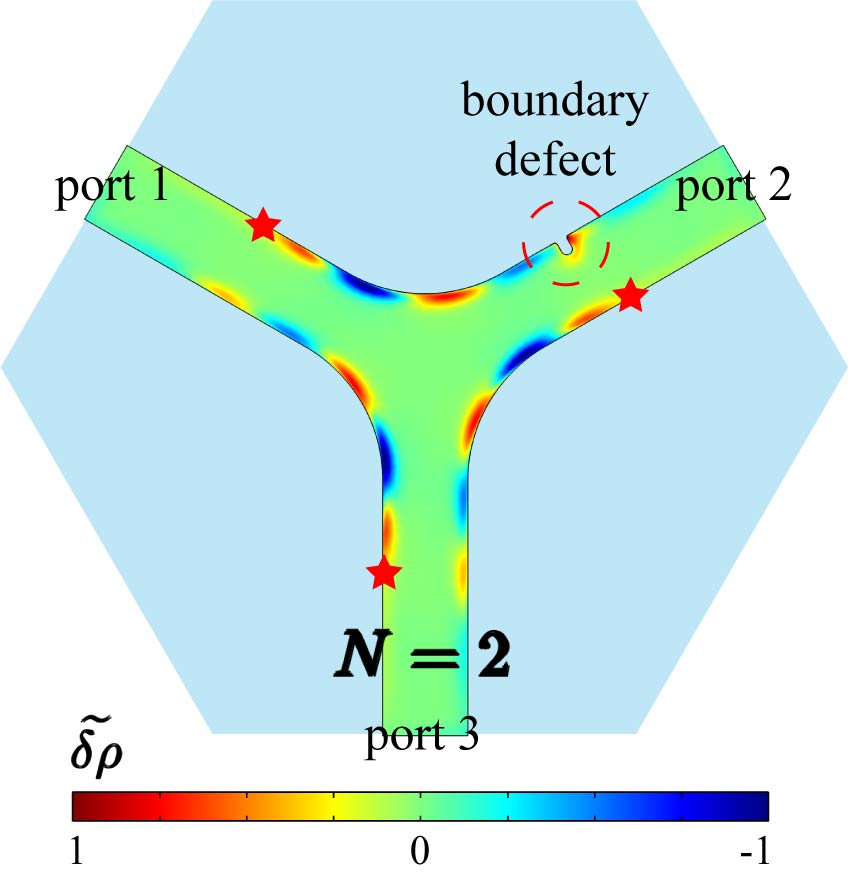}
    \caption{Time-domain simulations of the topological hydrodynamic circulator in the presence of large dissipation. Colorbar represents the charge density variation $\widetilde{\delta \rho}$. In this case, optical N-plasmons are still immune to back-scattering at the boundary defect.}
    \label{fig:fig6}
\end{figure}

In Fig.~\ref{fig:fig2}(a-c), we consider $\alpha=0.6817, \ \beta = 1.5263$ corresponding to the monolayer graphene experimental parameters in Table~\ref{tab:table1}. In Fig.~\ref{fig:fig2}(d-f), we consider $\alpha=3.9358, \ \beta = 1.5263$ for the $\mathcal{B}>1$ case. In Fig.~\ref{fig:fig2}(g-i), we consider $\alpha=0.6817i, \ \beta = 1.5263$ for the topologically trivial $N=0$ phase. In Fig.~\ref{fig:fig2}(j), we consider $\nu_H=0$  and other paramters are the same as the monolayer graphene parameters. In Fig.~\ref{fig:fig4}, we consider the monolayer graphene parameters, and $\beta$ is mainly determined by the neighbouring medium permittivity $\varepsilon$.

The solutions to $\delta \rho (x)$ of edge modes are:
\begin{equation}\label{drho}
    \delta \rho \, (x) \propto \Sigma_i (\tilde{q'_i}^2+\tilde{q}^2) \phi_i e^{-\eta_i \tilde{x}},
\end{equation}
where $\phi_i$ satisfies $\Sigma_j F_{ij} \phi_j =0$. 

From Eq.~(\ref{drho}), $\delta \rho \, (\tilde{x}=0) \propto \Sigma_i (\tilde{q'_i}^2+\tilde{q}^2) \phi_i$. In the low-loss limit, we can find that if there exists an edge mode satisfying all three kinds of electron fluid boundary conditions, $\delta \rho \, (\tilde{x}=0)  = 0$ since $\Sigma_i (\tilde{q'_i}^2+\tilde{q}^2) \phi_i$ is a linear combination of $\Sigma_i F_{3i}^0 \phi_i$ and $\Sigma_i F_{3i}^{+\infty} \phi_i$. Correspondingly, if a propagating edge mode satisfies $\delta \rho \, (\tilde{x}=0)  = 0$, then this mode can exist under all three kinds of electron fluid boundary conditions.

This can also be understood from the continuity equation only. Combining the Fourier transform of Eq.~(\ref{equation1b}) with respect to $x$ multiplied by $k$ (momentum corresponding to $x$),  $\delta \rho \, (\tilde{x}=0) \propto \lim_{|k|\to\infty} k f(k)$~\cite{cohen2018hall}, and the electron fluid boundary conditions, we can reach the same argument in the previous paragraph.

\section{Simulation details}\label{simulaton_details}
In this appendix, we present the graphene parameters employed in the topological hydrodynamic circulator simulations. We also provide a supplementary video for the time-domain simulations of optical N-plasmons in the supplementary materials. Fig.~\ref{fig:fig3}(b) corresponds to the simulation results at $\tilde{t}\approx 34$.

\begin{table}[h]
\caption{\label{tab:table1} Monolayer graphene parameters used in the simulations.}
\begin{tabularx}{\linewidth}{Xc}
\hline
Lattice constant, $a$ & $2.46~\si{\angstrom}$ \\
Electron density, $n_0$          & $2\times10^{12}~\mathrm{cm}^{-2}$   \\
Effective electron mass, $m$ & $ 0.0124m_e $ \\
Fermi velocity, $v_F$            & $1.1\times10^6~\mathrm{m/s}$                 \\

Biasing magnetic field, $B$     & $2~\mathrm{T}$                      \\

Cyclotron frequency, $\omega_c/2\pi$  & $4.52~\mathrm{THz}$ \\
Kinetic Hall viscosity, $\nu_H$         & $9.9\times 10^{-3}~\mathrm{m^2/s}$ \\
\hline
\end{tabularx}
\end{table}

\begin{figure}[t]
    \centering
    \includegraphics[width = 3.4in]{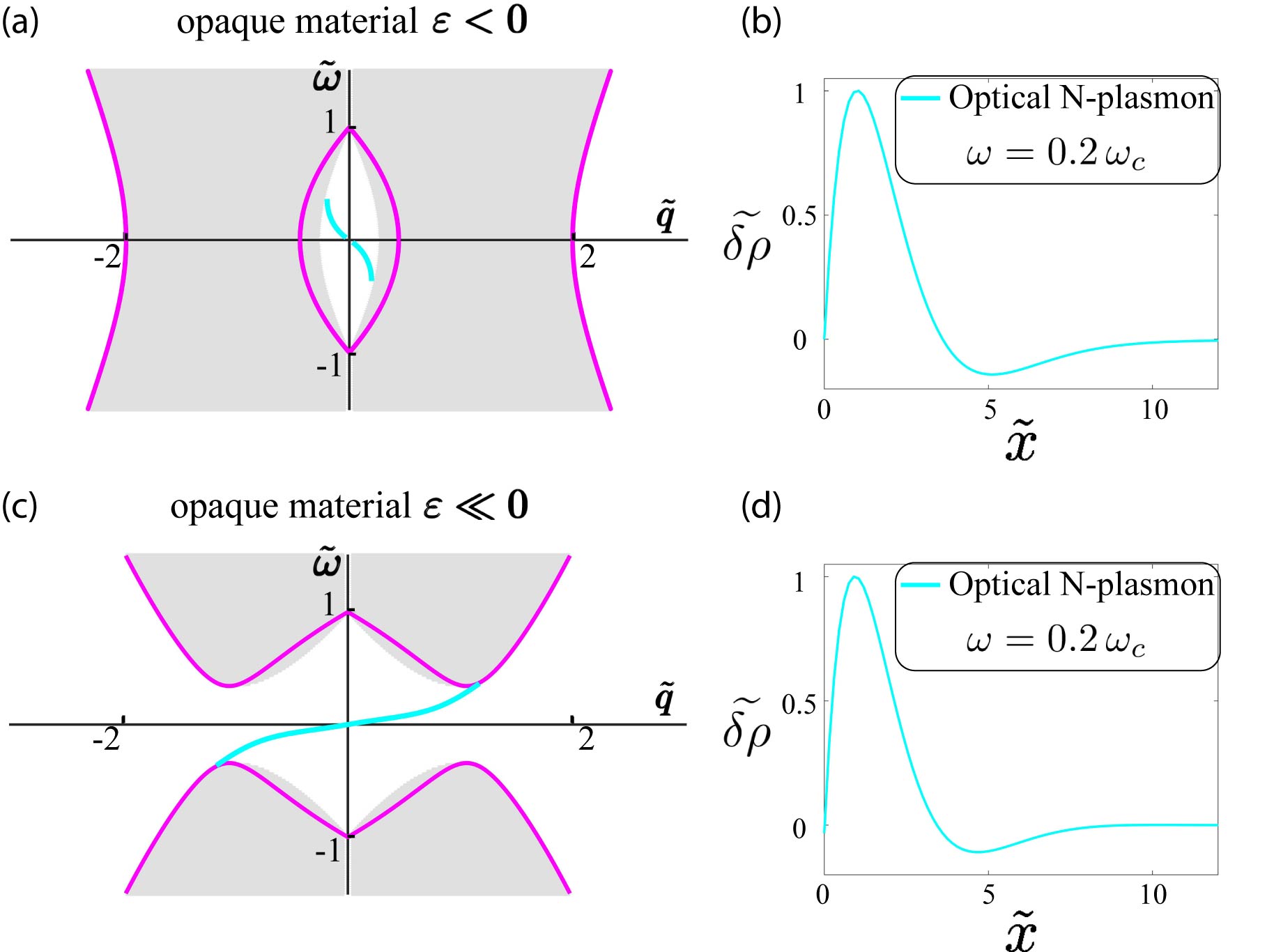}
    \caption{Optical N-plasmons near opaque materials with $\varepsilon < 0$. (a) Bulk bandgap at $q \neq 0$ can be closed when $|\varepsilon|$ is relatively small. In this case, the direction of optical N-plasmon group velocities will be reversed without breaking the topological properties. (b) With large $|\varepsilon|$, the bandgap at all $q$ will be opened. The group velocity of optical N-plasmons will have the same direction as the $\varepsilon>0$ case.}
    \label{fig:fig5}
\end{figure}

Here, we also demonstrate that the topological circulator will also have a robust performance in the presence of large dissipation. We consider the normal viscosity $\nu=9.9\times 10^{-4}~\mathrm{m^2/s}$ and damping rate $\tilde{\gamma}=\gamma / \omega_c = 0.0176$. In this case, we simulate the performance of the topological hydrodynamic circulator and show the results in Fig.~\ref{fig:fig6}. Boundaries of the circulator are taken to have finite slip length. Here, we can find that although the optical N-plasmons experience dissipation in the propagation process, it is still unidirectional and immune to back-scattering at the boundary defect in port 2. A video for the simulations of optical N-plasmons with large dissipation is also provided in the supplementary materials.

\section{Optical N-plasmons in the opaque surrounding medium}

In this appendix, we consider the dielectric materials surrounding the electron fluid in $N=2$ phase to be opaque with negative dielectric constant $\varepsilon$. Distinct from the transparent medium case where the bulk bandgap is always open for all $\tilde{q}$, the bulk bandgap can be closed at some $\tilde{q}$ points for negative $\varepsilon$ with small $|\varepsilon|$ values and reopened at all $\tilde{q}$ when $|\varepsilon|$ is large. We distinguish these two regimes by $\varepsilon<0$ and $\varepsilon \ll 0$. In Fig.~\ref{fig:fig5}(a,b), we show the dispersions of bulk magneto plasmons (magenta curve) and optical N-plasmons (cyan curve) and the profile of $\widetilde{\delta \rho}(\tilde{x})$ in the $\varepsilon<0$ case. It is worth noting that optical N-plasmons will persist in the bandgap close to $\tilde{q}=0$ points with a reversed direction. It is hard to excite the optical N-plasmons alone in this case since its frequency range is embedded in bulk bands. The reversal of the optical N-plasmon propagation direction is fundamentally different from the reversal of CEMP directions discussed in section~\ref{section2}, where bandgap closure never happens at any $\tilde{q}$ point. For the $\varepsilon \ll 0$ case, as is shown in Fig.~\ref{fig:fig5}(c,d), the bulk bandgap is reopened, and optical N-plasmons will have the same direction as in the $\varepsilon > 0$ case.

\clearpage
\bibliography{reference}

\end{document}